\documentclass{emulateapj}

\slugcomment{Submitted to the Astrophysical Journal}
\shorttitle{Photometric Redshift Errors}
\shortauthors{Quadri \& Williams}

\begin{document}

\title{Quantifying Photometric Redshift Errors in the Absence of
  Spectroscopic Redshifts}

\author{
Ryan F.~Quadri\altaffilmark{1},
Rik J.~Williams\altaffilmark{1,2},
}

\altaffiltext{1}{Leiden Observatory, Leiden University, NL-2300 RA,
  Leiden, Netherlands}
\altaffiltext{2}{Carnegie Observatories, Pasadena, CA 91101}
\email{quadri@strw.leidenuniv.nl}

\begin{abstract}

  Much of the science that is made possible by multiwavelength
  redshift surveys requires the use of photometric redshifts.  But as
  these surveys become more ambitious, and as we seek to perform
  increasingly accurate measurements, it becomes crucial to take
  proper account of the photometric redshift uncertainties.  Ideally
  the uncertainties can be directly measured using a comparison to
  spectroscopic redshifts, but this may yield misleading results since
  spectroscopic samples are frequently small and not representative of
  the parent photometric samples.  We present a simple and powerful
  empirical method to constrain photometric redshift uncertainties in
  the absence of spectroscopic redshifts.  Close pairs of galaxies on
  the sky have a significant probability of being physically
  associated, and therefore of lying at nearly the same redshift.  The
  difference in photometric redshifts in close pairs is therefore a
  measure of the redshift uncertainty.  Some observed close pairs will
  arise from chance projections along the line of sight, but it is
  straightforward to perform a statistical correction for this effect.
  We demonstrate the technique using both simulated data and actual
  observations, and discuss how its usefulness can be limited by the
  presence of systematic photometric redshift errors.  Finally, we use
  this technique to show how photometric redshift accuracy can depend
  on galaxy type.

\end{abstract}
\keywords{cosmology: observations -- galaxies: distances and redshifts
  -- methods: miscellaneous -- surveys}

\section{Introduction}
\label{sec:introduction}

Redshift surveys are a major and growing industry in astronomical
research.  The use of photometric, as opposed to spectroscopic,
redshifts in these surveys makes it possible to study a much larger
number of objects for a given amount of telescope time, and to study
the faintest sources.  However photometric redshifts are susceptible
to larger random and systematic errors, which can propagate into
derived quantities; in order to derive meaningful results using
photometric redshifts it is crucial to understand their uncertainties.
For instance both random and systematic redshift errors can lead to
systematic errors in the luminosity and mass functions
\citep{chen03,marchesini07}.  Studies of galaxy clustering can also be
strongly affected \citep{adelberger05,quadri08}.  In both of these
cases, it is possible to correct for the systematic errors in derived
quantities if the distribution of photometric redshift errors is
well-understood, but in practice this is seldom the case.  Surveys
that are designed to constrain the cosmological parameters require
especially tightly-constrained photometric redshifts, and significant
work has gone in to establishing the photometric redshift accuracy and
calibration requirements
\citep[e.g.][]{albrecht06,huterer06,mandelbaum08}.

The standard method used to estimate photometric redshift
uncertainties is to directly compare the photometric redshifts to the
spectroscopic redshifts for some subset of objects.  However
spectroscopic samples are frequently not representative of the full
photometric samples; at least at $z \gtrsim 1$, galaxies with
high-confidence spectroscopic redshifts are often brighter, bluer,
biased toward a specific sub-population (e.g. Lyman Break Galaxies or
AGN), or cover a different redshift range than the full photometric
sample.  Furthermore, if the parameters used to calculate photometric
redshifts are tuned to minimize the differences between the
photometric and spectroscopic redshifts, there is little guarantee
that these parameters are optimal for the full photometric sample.

The photometric redshift calculation itself also naturally produces an
estimate of the photometric redshift uncertainties.  For
template-fitting approaches \citep[e.g.][]{bolzonella00,brammer08},
the uncertainties are derived from the $\chi^2(z)$ of the template
fits.  However, in practice the uncertainties determined in this way
(not to mention the photometric redshifts themselves) can depend quite
sensitively on the shape and number of templates used.  Similarly, the
uncertainties derived when using empirical photometric redshift
algrorithms depend on the quality of the training set
\citep[e.g.][]{collister04}.

In this paper we describe a simple empirical method of using close
pairs of objects on the sky to estimate the width and shape of the
photometric redshift error distribution.  Because galaxies are
strongly clustered in real space, there is a high probability that any
one galaxy has nearby neighbors.  Therefore, close pairs of objects
will have a significant probability of lying at the same redshift, and
the differences in photometric redshifts of paired galaxies can then
be used to constrain the redshift errors.  We first illustrate the
method using a simulated data set, and then show examples using public
data.  Below we use the terms \emph{physical pairs} when referring to
objects that are physically-associated with each other (and thus lie
at similar redshifts), and \emph{projected pairs} when referring to
objects that lie at different redshifts.  Additionally, to avoid
confusion between an object's actual redshift and its photometric
redshift, we refer at times to the former quantity as its
\emph{spectroscopic redshift}.  All magnitudes are on the AB system.

\section{Method}
\label{sec:method}

\subsection{Overview}

Here we illustrate how it is possible to estimate the distribution of
photometric redshift errors in a completely empirical way, even in the
absence of spectroscopic redshifts.  The underlying principle is that
galaxies in an ordinary astronomical image will show significant
angular clustering (i.e.~an excess number of near neighbors over what
would be expected from a purely random distribution), which simply
reflects the real-space clustering projected on the sky.  But the
angular clustering arises only from galaxies that are physically
associated with each other, and thus lie at (nearly) the same
redshift.  In other words, a sample of all close pairs of objects in
an astronomical image will have a random contribution from projected
pairs, and an excess contribution from pairs in which both objects lie
at the same redshift.\footnote{Although we limit the discussion in
  this paper to close pairs, it is possible to use larger $N>2$
  associations of objects.}

\begin{figure}
  \epsscale{1.1}
  \plotone{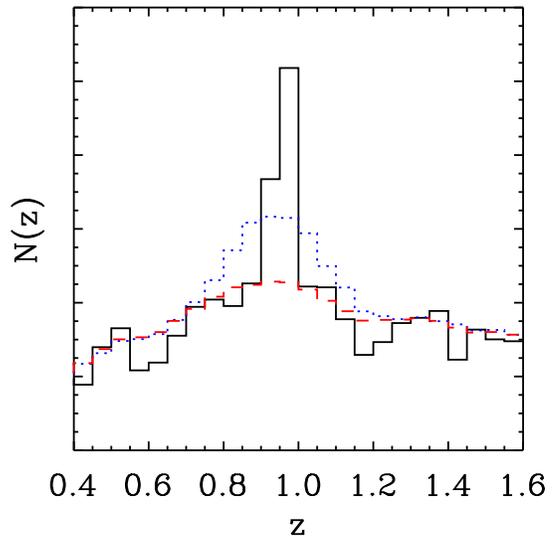}
  \caption{The redshift distribution of nearby neighbors of objects
    drawn from a lightcone based on the Millennium Simulation.  We
    select objects at $0.9<z<1.0$ and plot the redshift distribution
    of all neighbors (chosen, in this case, to have an angular
    separation $2.5\arcsec<\theta<15\arcsec$ from the central object)
    as the black histogram.  The spike shows that a significant number
    of the nearby neighbors lie at the same redshift as the central
    objects.  The blue dotted histogram has been calculated in the
    same way as the black histogram, except that the redshifts of all
    objects have been perturbed to simulate photometric redshifts; the
    spike is still visible, but has been broadened by the redshift
    errors.  The red dashed histogram shows the redshift distribution
    of neighbors after randomizing the galaxy positions, and simply
    reflects the overall redshift distribution of all objects in the
    catalog; this can be taken as an estimate of the contribution to
    $N(z)$ of projected close pairs.}
  \label{fig:Nz}
\end{figure}

To demonstrate this principle, we use mock observations generated from
the Millennium Simulation \citep{springel05}.  The method used to
create these ``lightcones'' is described by \citet{kitzbichler07}.  We
obtained the positions and redshifts of all simulated galaxies down to
$K = 23.9$ in a single $\sim 2 \rm{deg}^2$ lightcone from the
Millennium database \footnote{see http://www.g-vo.org/Millennium}.  We
select objects with $0.9<z<1.0$ in the lightcone, and determine the
redshift distribution of all objects lying within a small angular
separation of the selected objects.  This is shown by the black
histogram in Fig.~\ref{fig:Nz}.  The prominent spike at $0.9<z<1.0$
shows that many of these nearby neighbors lie at the same redshift.
We then create ``photometric redshifts'' for all objects in the
catalog by applying random Gaussian offsets to the true redshifts, and
repeat this procedure.  The blue dotted histogram shows the result;
the spike is still present, but has been broadened by the redshift
errors.  Finally, to estimate the contribution to $N(z)$ by close
pairs that arise only in projection, we randomize the angular
positions of objects in the catalog, and repeat the procedure again.
This randomization removes the clustering of sources, so now the only
pairs are projected pairs; the result is shown by the red dashed
histogram.  We can isolate the physical pairs, in a statistical sense,
by subtracting the red histogram from the blue, and can estimate the
distribution of photometric redshift errors from the width and shape of
the spike.

\subsection{Estimating Photometric Redshift Accuracy from
  Physically-Associated Pairs}

In the absence of spectroscopic redshifts, the dispersion of
photometric redshifts can be estimated by comparing the difference in
the photometric redshifts of the objects in physical pairs.  To see
how this is done, we model the photometric redshifts as being offset
from the true redshifts using

\begin{equation}
z_{phot} = \delta_z \cdot (1+z)+z,
\label{eq:1}
\end{equation}

\noindent where $\delta_z$ is a random deviate.  Eq.~\ref{eq:1}
implicitly assumes that the uncertainties are constant in units of
$(1+z)$ --- but note that this condition is at best only approximately
met in current data sets.  Since typical photometric redshift
uncertainties are significantly larger than the true redshift
differences in physically-associated pairs, we can assume that the
true redshifts of both objects in a pair are identical.

The best estimate of the true redshift of a pair is its mean
photometric redshift.  We can then measure the quantity

\begin{equation}
\Delta_z \equiv (z_{phot,1}-z_{phot,2})/(1+z_{mean}).
\label{eq:2}
\end{equation}

\noindent From eqs.~\ref{eq:1} and \ref{eq:2}, it can be shown that

\begin{equation}
\Delta_z \simeq \delta_1-\delta_2 - \frac{1}{2}(\delta_1^2-\delta_2^2),
\label{eq:3}
\end{equation}

\noindent where we have kept only the first- and second-order terms.
If $\delta_z$ follows a Gaussian distribution, and if $\delta_z << 1$,
then the dispersion in $\Delta_z$ is related to the dispersion in $\delta_z$
by

\begin{equation}
\sigma(\Delta_z) \simeq \sqrt2\sigma(\delta_z).
\label{eq:4}
\end{equation}

\noindent where we have additionally assumed that both objects in the
pair have similar uncertainties.  There may be times when it is useful
to consider close pairs of different types of objects, such as
bright-faint pairs, in which case this assumption will not hold and
the uncertainties should be added in quadrature.

\subsection{Subtracting out the Projected Pairs}
\label{sec:contaminants}

Unless pairs with only very small angular separations are used, the
number of projected pairs will be comparable to, or significantly
greater than, the number of physical pairs.  It then becomes necessary
to statistically subtract out the contaminants.  The expected number
and distribution of contaminants can be easily estimated by
randomizing the positions of the galaxies from which the observed
pairs are drawn (while keeping the redshifts the same), and by
detecting the random pairs.  The random positions should follow the
same observing geometry constraints as the observed positions
(i.e.~avoiding the locations of bright stars or other image
artifacts), and the process can be repeated several times to reduce
the uncertainty.

\begin{figure*}
  \epsscale{1.1}
  \plotone{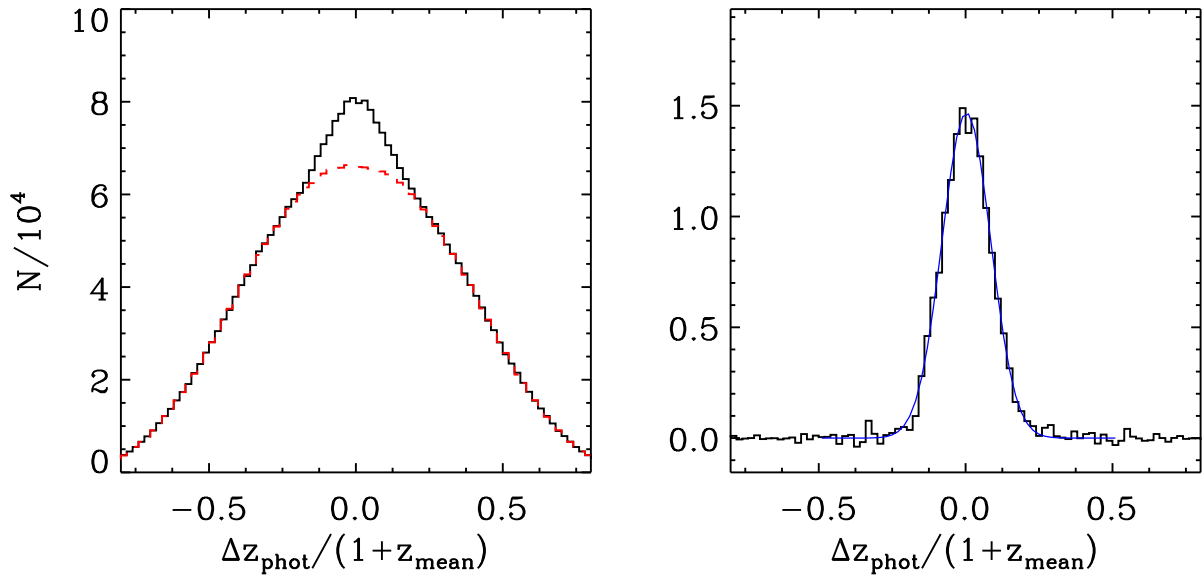}
  \caption{\emph{Left:} The black histogram shows the distribution of
    $\Delta z_{phot}/(1+z_{mean})$ for galaxy pairs selected with
    $1<z_{phot}<2$ in a simulated lightcone.  The red histogram shows
    the same thing, but drawn from a catalog in which the galaxy
    positions have been randomized in order to estimate the
    distribution for chance projections.  \emph{Right:} The histogram
    shows an estimate of the distribution of redshift separations for
    physical galaxy pairs, which comes from subtracting the red
    histogram from the black histogram in the left panel.  The blue
    curve shows the best-fitting Gaussian, which has a width that is
    greater than the photometric redshift errors by a factor of $\simeq
    \sqrt2$.}
  \label{fig:lcone}
\end{figure*}

For purposes of illustration, we create ``photometric redshifts'' for
objects in the lightcone by perturbing the true redshifts with
Gaussian random deviates with $\sigma = 0.06$, which is a typical
error for high signal-to-noise objects in high-quality data sets at
$z>1$.  We select objects in the lightcone with $1<z_{phot}<2$, and
identify all pairs with an angular separation 2.5--15$\arcsec$.  The
lower limit is applied to minimize the effect of blending on the
object photometry (this is obviously not an issue for the simulated
data used in this section, but will be an issue for actual data).  The
upper limit was chosen arbitrarily; a larger value would yield more
pairs, and thus a more accurate estimate of the redshift
uncertainties, but 15$\arcsec$ is sufficient for our purposes and
limits the computational expense.  The left panel of
Fig.~\ref{fig:lcone} shows the distribution of $\Delta
z_{phot}/(1+z_{mean})$ for both the observed pairs and the pairs found
in the randomized catalog.  In the right panel we subtract the
randomized histogram from the true histogram; this subtraction is a
statistical correction for the projected pairs.  Also shown is the
best-fitting Gaussian, which has width $\sigma = 0.084$, which is
essentially identical to the expected value of $\sqrt2 \cdot 0.06$
from eq.~\ref{eq:4}.\footnote{Note that each pair where both objects
  lie at $1<z_{phot}<2$ is counted twice, whereas pairs where only one
  object lies in the redshift range is counted once.}
\footnote{Although it would simplify the analysis somewhat to select
  pairs where both objects lie within a given redshift range, this can
  yield an artificial reduction in the photometric redshift scatter.
  For example, in the extreme case of a very narrow redshift selection
  window, the requirement that both objects lie in this window would
  mean that both objects have essentially identical photometric
  redshifts, leading to the conclusion that the photometric redshifts
  are extremely accurate.}.

It can be numerically intensive to generate several random catalogs
based on the actual data, and to identify all close pairs, as
described above --- especially if one wishes to repeat the
measurements many times for different samples of galaxies.  In
practice we speed up the method by creating a single large catalog of
points with random angular positions, detecting the pairs, and storing
the vector of pair separations.  Then we can quickly mimic the random
catalogs described above for any range of angular separations by
assigning to each value of the separation a value of $\Delta_z$, where
the photometric redshifts in eq.~\ref{eq:2} are drawn at random from
the data.  The remaining step is to scale the number of random pairs
to the number there would be if the random catalog had the same number
of objects as the data catalog.

\subsection{Non-Gaussian Errors}

In the previous sections we dealt with the case of Gaussian
photometric redshift errors.  But in actual data sets the average
photometric redshift probability distribution will not generally be a
perfect Gaussian, and hence $\Delta_z$ will also deviate from a
Gaussian.  It is therefore of interest to consider other functional
forms, particularly those with more prominent wings than a simple
Gaussian.  One possibility is to consider the case of error
distributions that are the sum of two Gaussians.  We first note that
the distribution of $z_{phot,1}-z_{phot,2}$ is simply the error
distribution convolved with itself \footnote{Rigorously speaking, the
  distribution is cross-correlated with itself rather than convolved
  with itself, but the two operations are equivalent for the even
  functions considered here.}.  A single Gaussian convolved with
itself will become broader by a factor of $\sqrt2$, which explains
presence of that factor in eq.~\ref{eq:4}.  A double Gaussian
convolved with itself results in a triple Gaussian (i.e.~each of the
two Gaussians convolved with themselves, plus a third Gaussian which
is the two Gaussians convolved with each other).

To illustrate how this works, we produce ``photometric redshifts'' in
the lightcone by perturbing the spectroscopic redshifts by a double
Gaussian, with widths $\sigma_1=0.03$ and $\sigma_2=0.09$, and we set
the relative areas of the second to the first Gaussian to $r=0.5$.
This is the same as saying that $2/3$ of the objects have a redshift
error given by $\sigma_1$ and $1/3$ have an error given by $\sigma_2$.
We then calculate $\Delta(z_{phot})/(1+z_{mean})$ for the close pairs,
and fit a function of the form

\begin{equation}
  F(x) = AG(x,2\sigma_1^2) + A2rG(x,\sigma_1^2+\sigma_2^2) + Ar^2G(x,2\sigma_2^2)
\label{eq:5}
\end{equation}

\noindent where $G(x,\sigma^2)$ is a normalized Gaussian with variance
$\sigma^2$, and $A$ is just an overall normalization factor.  Thus the
fitting parameters are $(A, \sigma_1, \sigma_2, r)$, and we have
increased the number of parameters relative to the single Gaussian
case by two.

\begin{figure}
  \epsscale{1.1}
  \plotone{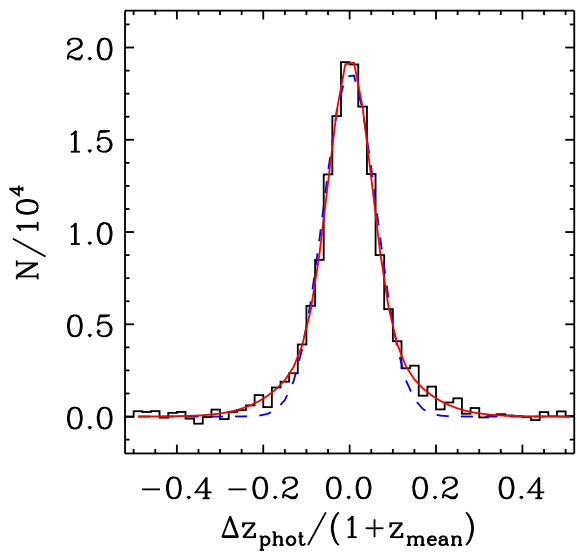}
  \caption{The distribution of $\Delta z_{phot} /(1+z_{mean})$ after
    correcting for projected pairs.  The photometric redshift errors
    have been parameterized as the sum of two Gaussians, so the
    appropriate fitting function for $\Delta z_{phot} /(1+z_{mean})$
    is the sum of three Gaussians (see text).  This fit is shown by
    the red curve.  The dashed blue curve shows the result of a single
    Gaussian fit.}
\label{fig:triplegauss}
\end{figure}

Figure \ref{fig:triplegauss} shows the result.  We obtain $(\sigma_1,
\sigma_2, r) = (0.031,0.083,0.50)$, which is close to the input
values.  This figure also shows the result of fitting a single
Gaussian to the distribution; the fit is obviously not as good, but
still give a reasonable estimate of the errors, with $\sigma = 0.06$.
In practice, fits using eq.~\ref{eq:5} can become somewhat unstable in
certain regimes of parameter space due to covariance in the fitting
parameters, or due to poor signal-to-noise.  It is useful to constrain
the fitting parameters so they do not reach very small, or negative,
values.

Another formula that is useful to parameterize photometric
redshift errors in the case of non-Gaussian tails is the Lorentz
distribution, $L(x,\gamma)$.  Convolving a Lorentz distribution with
itself results in another Lorentz distribution where $\gamma$ is
increased by a factor of 2 \citep{dwass85}.  So in this case the
fitting function for $\Delta z_{phot} /(1+z_{mean})$ would be

\begin{equation}
F(x) = AL(x,2\gamma)
\label{eq:6}
\end{equation}

where the fitting parameters are $(A,\gamma)$.

\subsection{Estimating the Catastrophic Failure Rate}

Thus far we have modeled the photometric redshift errors as small
perturbations on the true redshifts.  However, actual photometric
redshifts are subject to ``catastrophic failures.''  These outliers
may be caused by multiple minima in $\chi^2(z)$, or may arise from a
mismatch between the observed galaxy spectral energy distributions and
the galaxy templates used when calculating photometric redshifts.  It
is possible to estimate the outlier rate by looking for an excess of
neighbors with widely discrepant photometric redshifts \citep[see
also][who use the cross-correlation between galaxies in different
photometric redshift bins to detect the existence of catastrophic
failures]{erben09}.

\begin{figure}
  \epsscale{1.1}
  \plotone{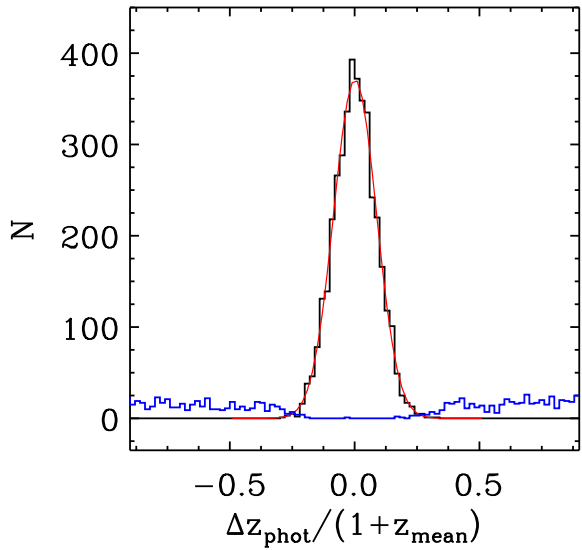}
  \caption{The effect of catastrophic redshift failures.  The black
    histogram shows the distribution of $\Delta z_{phot}
    /(1+z_{mean})$ for all physical galaxy pairs which have Gaussian
    photometric redshift errors in the simulations, while the blue
    histogram shows the contribution from pairs that contain at least
    one catastrophic failure.  In this example the catastrophic
    outlier rate is $30\%$.}
  \label{fig:catastrophics}
\end{figure}

To demonstrate the effects of catastrophic failures, we first detect
only the physical pairs in the lightcone.  We then generate
``photometric redshifts'' for each object as described in
\S~\ref{sec:contaminants}, but now include outliers by assigning random
redshifts that are constrained to be at least $5\sigma$ away from the
true redshift for $30\%$ of the objects (the catastrophic failure rate
is lower than this for normal galaxies in current multiwavelength
surveys, but we exaggerate the effect to provide a clear
illustration).  Figure \ref{fig:catastrophics} shows the result.  The
best-fitting Gaussian still has a width of $\sim \sqrt2\cdot0.06$, but
the distribution of $\Delta z_{phot}/(1+z_{mean})$ has broad wings due
to the outliers.

The outlier rate can be estimated from the area under the histogram at
large $|\Delta z_{phot}|/(1+z_{mean})$.  In this case, the fractional area
at $|\Delta z_{phot}|/(1+z_{mean}) > 0.3$ is $30.4\%$, which is close to the
expected value of $30\%$.

We note that the extended, flat wings shown in Figure
\ref{fig:catastrophics} are due to the flat redshift distribution that
we use for the redshift outliers.  In actual data sets, the outliers
will frequently not have a flat redshift distribution, for instance
because of the double redshift solutions which can be caused by
confusion between the Lyman and Balmer/4000\AA\ breaks.  Nevertheless,
this example illustrates that even the central part of the
distribution of $\Delta z_{phot} /(1+z_{mean})$ can be affected by the
outliers (where the blue and red histograms overlap in Figure
\ref{fig:catastrophics}), so it is important to fit a Gaussian only
over the regions that are clearly dominated by non-outliers.  The
location of this region will depend on what is chosen to constitute a
catastrophic failure.  In practice, the catastrophic failure rate can
not always be tightly constrained, and relies on accurate modeling of
the projected pairs.

\section{The Effects of Systematic Errors in the Photometric Redshifts}
\label{sec:cosmos}

In the previous section we demonstrated the close-pairs technique on
mock data.  Here we use the technique to determine the photometric
redshift errors in an actual dataset, and discuss the effects of
systematic redshift errors.

The Cosmic Evolution Survey \citep[COSMOS;][]{scoville07} is a 2
deg$^2$ multiwavelength survey, and was conducted with the primary
goal of studying the relationship between galaxy evolution and
large-scale structure.  A unique aspect of this survey is the number
of observed filters, with 30 bands from the ultra-violet to the
mid-infrared.  Particularly valuable is the deep medium-band optical
imaging, which traces galaxy spectral energy distributions with much
higher resolution than is possible with standard broadband filters.
\citet{ilbert09} present photometric redshifts for the COSMOS field.
They use a template-fitting approach to derive the photometric
redshifts, paying particular attention to the choice of templates and
to the effects of emission lines.  The medium-band imaging allows
\citet{ilbert09} to achieve extremely accurate photometric redshifts
out to $z \sim 1$; for the brightest sources, with $I<22.5$, they
quote a typical error in $|\Delta(z)|/(1+z)$ of 0.007.

Another unique aspect of this field is the large number of
spectroscopic redshifts available from the zCOSMOS survey
\citep{lilly09}.  For the purposes of this paper, these spectroscopic
redshifts are extremely useful as the zCOSMOS-bright sample has a high
level of completeness for $I<22.5$, and the objects with secure
spectroscopic redshifts are a relatively unbiased subset of the parent
population.  Thus we can use these spectroscopic redshifts to obtain
an independent test of the photometric redshift errors.  In what
follows, we reject objects classified as stars or x-ray sources in the
photometric catalog.  We also only make use of the secure
spectroscopic redshifts (with confidence class 3 or 4).

\begin{figure*}
  \epsscale{1.1}
  \plotone{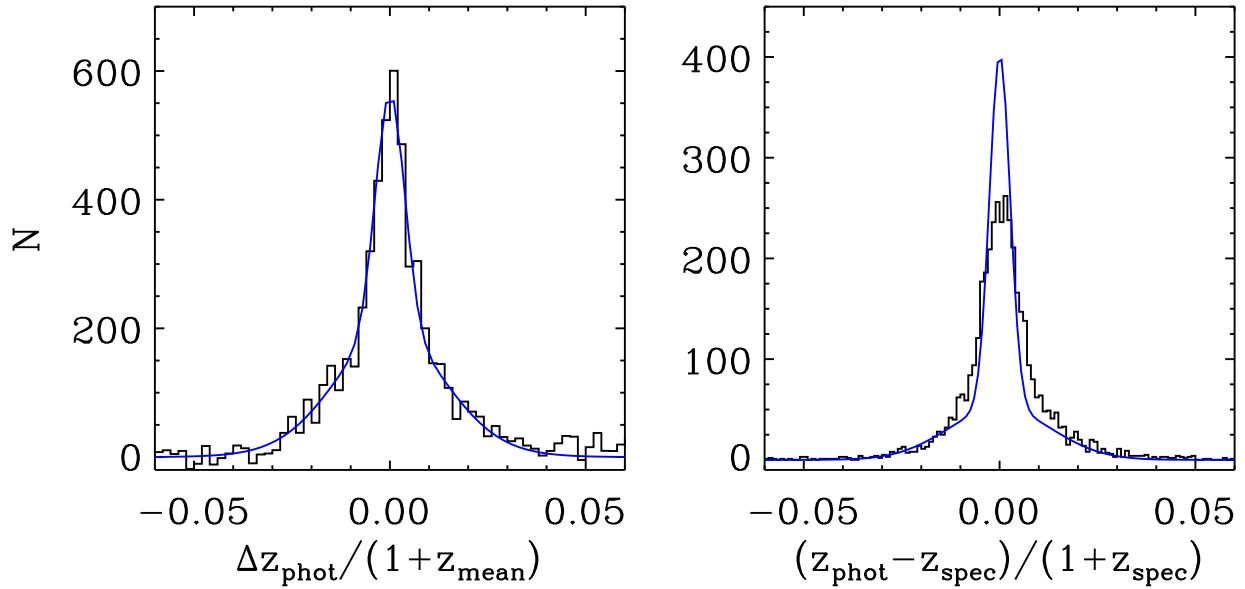}
  \caption{\emph{Left}: The distribution of $\Delta z_{phot}
    /(1+z_{mean})$ for galaxies with $17.5<I<22.5$ in the COSMOS
    field.  The blue curve is a fit.  \emph{Right}: The distribution
    of $(z_{phot}-z_{spec})/(1+z_{spec})$ for objects with
    spectroscopic redshifts from zCOSMOS, again for galaxies with
    $17.5<I<22.5$.  The curve is the error distribution that would be
    inferred from the fit in the left panel.  Although the typical
    size of the errors is recovered well to first order, the shape of
    the error distribution is not perfect.}
\label{fig:cosmos_bright}
\end{figure*}

To demonstrate how well we can recover the photometric redshift errors
from real data using the technique described in this paper, we begin
by comparing our estimate of the errors to a direct measurement of the
errors made by comparing photometric and spectroscopic redshifts for
individual galaxies.  For this we use the zCOSMOS-bright sample.  The
left panel in Figure \ref{fig:cosmos_bright} shows the $\Delta
z_{phot} /(1+z_{mean})$ for galaxies with $17.5<I<22.5$.  The blue
curve shows the best fit, using the fitting function eq. \ref{eq:5}.
The right panel shows the direct measurement of the photometric
redshift errors in a standard $(z_{phot}-z_{spec})/(1+z_{spec})$ plot,
and the blue curve shows the predicted errors from the fit in the left
panel.  Although to first order we do recover the typical magnitude of
the errors reasonably well, the distribution of errors is obviously
not perfect as the central region is too strongly peaked.

\begin{figure*}
  \epsscale{1.1}
  \plotone{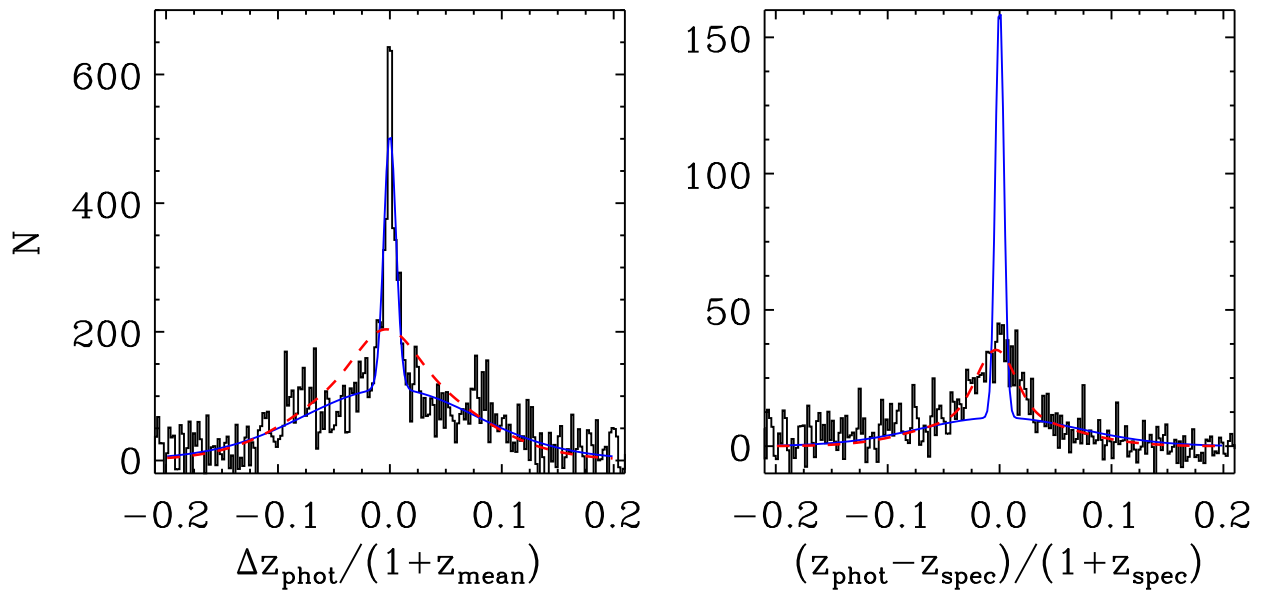}
  \caption{\emph{Left}: The distribution of $\Delta z_{phot}
    /(1+z_{mean})$ for galaxies with $24<I<25$ in the COSMOS field.
    The solid blue curve is a fit.  \emph{Right}: The distribution of
    $(z_{phot}-z_{spec})/(1+z_{spec})$ for pairs of objects with one
    photometric and one spectroscopic redshift (see text).  The solid
    blue curve is the error distribution that would be inferred
    from the fit in the left panel.  The dashed red curve is a fit to
    the data in this panel, and the dashed red curve in the left
    panel is the corresponding prediction for $\Delta z_{phot}
    /(1+z_{mean})$.  The differences between the red and blue curves
    in both of these panels illustrates the effect of photometric
    redshift ``attractors,'' which is substantial for faint galaxies
    in the COSMOS photometric redshift catalog.}
\label{fig:cosmos_fainter}
\end{figure*}

This effect can be seen more strongly at fainter magnitudes, as
demonstrated in Figure \ref{fig:cosmos_fainter}.  The left panel shows
$\Delta z_{phot} /(1+z_{mean})$  for galaxies with $24<I<25$.  Here the error
distribution appears highly non-Gaussian, with broad tails and a very
narrow peak.  In this case we cannot directly measure the photometric
redshift errors as done in Figure \ref{fig:cosmos_bright} since
there are few spectroscopic redshifts available for galaxies at these
faint magnitudes.  Thus we follow a different approach of estimating
the error distribution from \emph{photometric-spectroscopic} pairs,
i.e.~we measure the redshift separation of close pairs where one
object has a spectroscopic redshift from the zCOSMOS-bright sample and
the other object has $24<I<25$.  In this case, the projected pairs can
be subtracted out in a manner analogous to that described in \S
\ref{sec:contaminants}, except that the positions of
spectroscopic sample are fixed and only the positions of the
photometric sample are randomized.  The photometric redshift errors
estimated in this way are shown in the right panel of Figure
\ref{fig:cosmos_fainter}.  Again, the solid blue curve shows the error
distribution that would have been predicted from the fit in the left
panel.  Obviously, the strong central peak is an artifact and does not
represent the true errors.  The red dashed curve in this panel is a fit
using a double-Gaussian fitting function, and the red curve in the
left panel shows the resulting prediction for
$\Delta z_{phot} /(1+z_{mean})$ .

The fact that many close pairs of objects have photometric redshifts
that are closer than expected based on the true errors is largely due
to an artifact of the photometric redshift algorithm.  It is a common
feature of many data sets that there are artificial spikes in the
photometric redshift distribution.  These spikes may result from the
particular filter/template combination, or may be due to systematic
errors in object colors.  For instance, if the observed galaxy colors
in two closely-spaced filters are systematically too red --- due to
poor PSF matching or zeropoint errors --- then the photometric
redshift code may interpret the red colors as being due to 4000\AA\
breaks in the galaxy SEDs, with the effect that many objects will have
artificially similar photometric redshifts.

This illustrates the fundamental limitation of the technique described
in this paper, which is that its usefulness is reduced in the case of
significant systematic redshift errors.  A more frequently-discussed
type of systematic error, in which all photometric redshifts are over-
or underestimated, will go completely undetected using the method
described in \S \ref{sec:method} \citep[however in some cases such
biases are relatively unimportant, so long as they are
small;][]{quadri07}.  Artificial spikes in the photometric redshift
distribution are a somewhat different type of systematic error; this
type of error is not nearly as noticeable in most data sets as it is
in Figure \ref{fig:cosmos_fainter}, and we have made use of the COSMOS
field here primarily for its illustrative value.  This does not
necessarily mean that the COSMOS photometric redshifts suffer from
redshift ``attractors'' --- also sometimes called ``redshift
focusing'' --- more than other photometric redshift catalogs; it may
simply mean that the random errors are so small in this case that the
systematic errors become important.

\section{Differential Photometric Redshift Errors}
\label{sec:differential_errors}

In this section we investigate how photometric redshift errors depend
on redshift, signal-to-noise ratio (S/N), and galaxy type.  Because in
most data sets the photometric redshifts are constrained primarily by
the locations of Lyman break and/or the Balmer/4000\AA\ break, objects
with weak or undetected breaks will have comparatively uncertain
photometric redshifts.  It is therefore expected that redshift
accuracy will depend not just on galaxy brightness, but also on galaxy
type.

Here we use public data in the field observed by the UKIDSS Ultra-Deep
Survey \citep[UDS;][]{lawrence07,warren07}.  We use an updated version
of the UDS catalog that was presented by \citep{williams09}, and
details of the data, photometry, and redshifts can be found in that
work.  Briefly, this catalog includes near-infrared (NIR) imaging from
the UDS, optical imaging from the Subaru-XMM Deep Survey
\citep[SXDS;][]{furusawa08}, and infrared imaging from the
\emph{Spitzer} Wide-Area Infrared Extragalactic Survey
\citep[SWIRE;][]{lonsdale03}.  The field size with complete
multiwavelength coverage is $\sim$0.65$\rm{deg}^2$.  The latest
version of our catalog includes $H$-band imaging in the NIR from the
UDS data release 3 and $V$-band imaging from the SXDS data release 1.
We have also added $u^*$-band imaging from the Canada-France Hawaii
Telescope (CFHT) that was taken as part of Program ID \#07BC25
(P.I.~O.~Almaini), and was downloaded from the CFHT archive.  Those
data were kindly reduced for us by H.~Hildebrandt using the procedures
described in \citet{erben09} and \citet{hildebrandt09}.  Thus the
updated catalog has complete $u^*BVRi'z'JHK3.6\mu\rm{m}4.5\mu\rm{m}$
photometry.

The photometric redshifts were calculated from the updated catalog
using the EAZY code \citep{brammer08}.  We did not perform any tuning
of the default EAZY parameters, with the single exception of reducing
the amplitude of the template error function to 0.5, which has been
found to provide better results in several different data sets
(G.~Brammer, private communication).  Additionally we use an updated
template set with a new treatment of emission lines (Brammer et al.,
in prep.).

To illustrate the effect of galaxy type on redshift accuracy, we
separate galaxies into star-forming and quiescent populations
according to the bimodality in a rest-frame $U-V$ vs.~$V-J$
color-color diagram \citep{williams09}.  We limit the sample to
$K<22.9$, and reject galaxies with high $\chi^2$ values from the
template fits as those objects tend to have very inaccurate
photometric redshifts and are frequently AGN.

We estimate accuracy in $\Delta(z)/(1+z)$ in four redshift bins:
$0.3<z_{phot}<0.7$, $0.7<z_{phot}<1.2$, $1.2<z_{phot}<1.7$, and
$1.7<z_{phot}<2.2$.  Within each redshift bin, we separate galaxies
into a bright and faint subsample according to the median
signal-to-noise ratio (S/N) of all galaxies in that bin.  Although
most studies classify galaxies according to the S/N in the detection
band, it is not obvious that this is a relevant statistic.  In this
case, the detection band is $K$, which does not actually play a major
role in constraining the redshifts at $z<2$.  Since photometric
redshifts are most strongly constrained at these redshifts by the
identification of the 4000\AA\ break, we use the S/N in the closest
band redward of the break.  Thus an old galaxy with a strong break,
which may have high S/N redward of the break and low S/N blueward of
the break, will be appropriately classified as high S/N since the
location of the break will be tightly-constrained.

\begin{figure*}
  \epsscale{1.1}
  \plotone{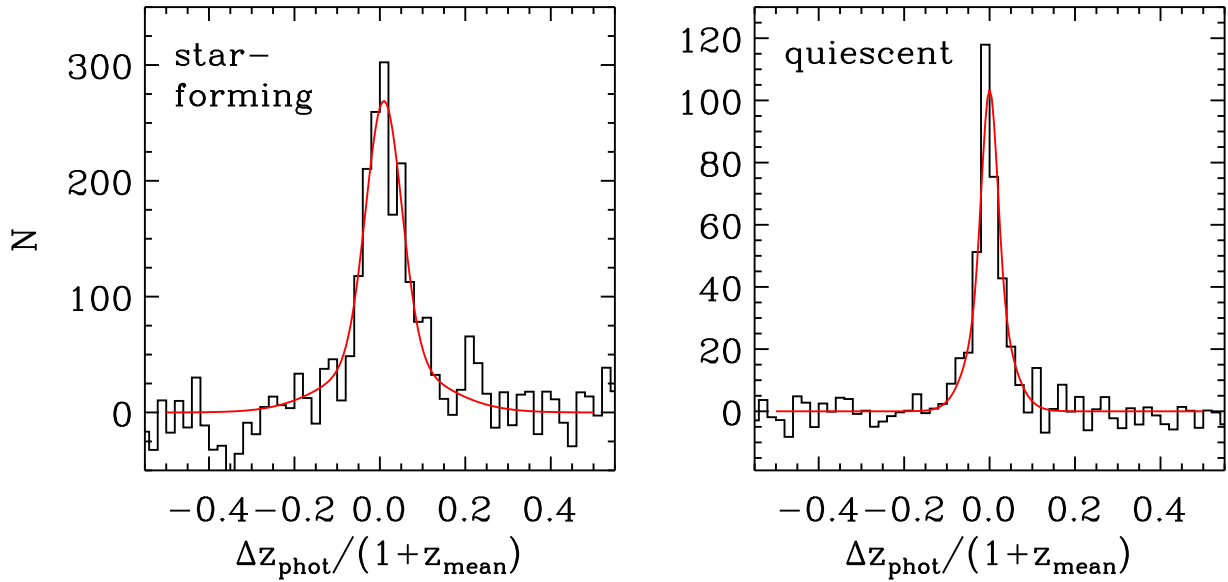}
  \caption{The distribution of $\Delta z_{phot}/(1+z_{mean})$ for
    star-forming galaxies and quiescent galaxies at $1.2<z_{phot}<1.7$
    in the UDS.  The quiescent galaxies have more accurate photometric
    redshifts.}
\label{fig:uds}
\end{figure*}

\begin{figure}
  \epsscale{1.1}
  \plotone{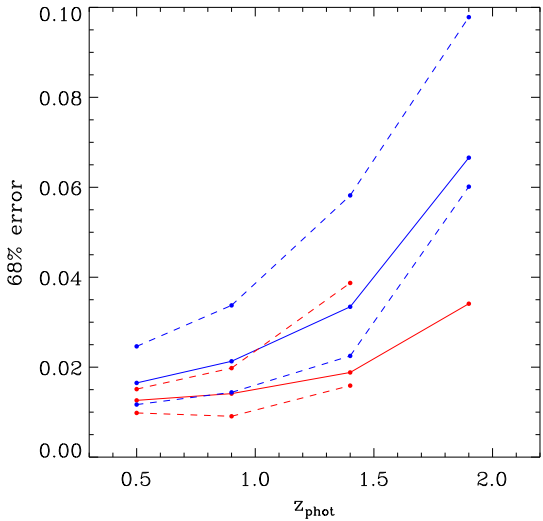}
  \caption{The $68\%$ errors in $\Delta(z)/(1+z)$ as function of
    redshift for star-forming (blue curves) and quiescent (red curves)
    galaxies.  The solid curves are for the full sample of $K<22.9$
    objects, while the lower dashed curves are for bright galaxies and
    the upper dashed curves are for faint galaxies.  A galaxy is
    classified as bright (faint) if the S/N in the band immediately
    redward of the 4000\AA\ break is higher (lower) than the median
    for all galaxies in that redshift bin.}
\label{fig:differential_accuracy}
\end{figure}

For each galaxy sample, we fit the distribution of $\Delta z_{phot}
/(1+z_{mean})$ using eq.~\ref{eq:5}.  Figure \ref{fig:uds} shows the
result in the $1.2<z_{phot}<1.7$ redshift bin (without splitting the
samples by S/N).  It is immediately apparent that the quiescent
galaxies have more accurate redshifts than the star-forming galaxies.
This is more clearly demonstrated in Figure
\ref{fig:differential_accuracy}, which shows the $68\%$ uncertainty in
$\Delta(z)/(1+z)$ that we estimate by integrating the inferred
photometric redshift error distribution.  The red and blue solid curve
show how this quantity changes with redshift for the quiescent and
star-forming galaxies, respectively.  The upper and lower dashed
curves show the accuracy for the fainter and brighter subsamples of
each population.

The quality of the photometric redshifts in the UDS is impressively
good.  The quiescent galaxies, in particular, have extremely accurate
redshifts at $z \lesssim 1$.  This is confirmed by a direct comparison
of photometric with spectroscopic redshifts using the substantial
spectroscopic sample of $z \sim 1$ passively-evolving galaxies from
\citet{yamada05}; this yields $\Delta(z)/(1+z) = 0.015$, in good
agreement with the results shown in Figure
\ref{fig:differential_accuracy}.

The quiescent galaxies do show significantly more accurate photometric
redshifts than the star-forming galaxies over all redshifts probed
here.  For certain types of studies, such differential photometric
redshift errors can adversely affect the results.  For instance, the
increased errors for star-forming galaxies can lower the inferred
correlation length \citep{quadri07}, leading to an artificial
trend of clustering with star formation properties.  Another example
is the mass/luminosity function: redshift errors will tend to flatten
these functions relative to their true values (\citealt{chen03}; but
see \citealt{marchesini07}), and can lead to an artificial difference
between these functions for star-forming and quiescent galaxies.

The result that quiescent galaxies have more accurate redshifts is
obviously somewhat dependent on image depth and filter coverage; our
deep images and closely-spaced optical and NIR filters allow us to
pinpoint the location of the Balmer/4000\AA\ break for quiescent
galaxies quite accurately, while the lack of ultra-violet imaging
means that we cannot detect the Lyman break of star-forming galaxies
at these redshifts.  It is entirely possible that with different data
the star-forming galaxies would have photometric redshift accuracy
comparable to, or even better than, the quiescent galaxies.

\section{Summary and Discussion}
\label{sec:summary}

The use of photometric, as opposed to spectroscopic, redshifts makes it
possible to study a much larger number of objects for a given amount
of telescope time.  But photometric redshift errors will propagate
through many different types of analyses, and in practice may comprise
a significant source of error in derived quantities.  For this reason
a realistic estimate of the distribution of photometric redshift
errors is necessary.  Obtaining large, representative samples of
spectroscopic redshifts with which to directly measure the photometric
redshift errors is observationally expensive, and often completely
unfeasible.  For this reason it is of great interest to have a method
to estimate the size and distribution of such errors that can be
applied with limited, or even non-existent, spectroscopic samples.

In this paper we have presented such a method.  It is based on the
idea that a close association of two or more galaxies on the sky may
represent a true physical association, in which case the objects will
lie at nearly the same redshift and the differences between their
photometric redshifts constrains the typical errors.  We have
described a simple implementation of this idea that makes use of close
galaxy pairs, where the best estimate of the true redshift of a pair
is taken to be the mean of the photometric redshifts.  We have
described how to estimate the photometric redshift error distribution
from the difference in photometric redshifts, as well as how to
estimate the catastrophic failure rate.  This technique requires
applying a statistical correction for pairs that arise from chance
projections along the line of sight, and this is easily done by
randomizing the galaxy positions and repeating the analysis.  Although
in this paper we have focused on the redshift range $0.5 \lesssim z
\lesssim 2$, the basic technique can be applied at both significantly
lower and higher redshifts.

The concept of using angular associations between galaxies to
constrain their redshifts is not entirely new.  \citet{newman08} uses
a cross-correlation between a spectroscopic and a photometric sample
of galaxies to infer the true redshift distribution of the photometric
sample.  \citet{erben09} uses the cross-correlation between galaxies
selected in two disjoint photometric redshift bins to quantify the
photometric redshift errors.  \citet{kovac09} modifies the photometric
redshift probability distribution of objects on the basis of the
spectroscopic redshifts of nearby objects.  The technique presented in
this paper represents a significant step forward because it is simple
to implement, the results are easy to interpert, and it can be applied
with limited (or even with a complete lack of) spectroscopic
information.

As a first application of our method, we have shown that quiescent
galaxies will on average have more accurate photometric redshifts than
star-forming galaxies in broadband optical/NIR surveys out to at least
$z \sim 2$.  This is because quiescent galaxies have a strong break in
their SEDs near 4000\AA, and if the location of this break can be
pinpointed using the observed photometry, the redshift will be tightly
constrained.  Star-forming galaxies, on the other hand, have weaker
features in their SEDs over the range of observed wavelengths.
Differential photometric redshift errors can lead to differential
effects in derived quantities, such as luminosity or mass functions,
and should be taken into account when comparing such quantities
between samples.

A significant limitation of the method presented in this paper arises
from systematic errors in the photometric redshifts.  One type of
systematic error, in which all photometric redshifts are biased in one
direction, will go completely undetected.  However if a particular
class of galaxies (e.g. galaxies on the red sequence) is subject to
such a bias, whereas another class (in the blue cloud) is not, then
this bias will become apparent by looking at cross-pairs (red-blue pairs).
Another type of systematic error is when the photometric redshift
distribution shows artificial spikes.  This is particularly
problematic, as it means that the photometric redshifts of a pair of
objects may both be drawn into the spike, leading to a smaller
relative redshift difference and an underestimate of the true redshift
errors.  In extreme cases, when this type of error is comparable to
the random errors, this effect can lead to highly disturbed error
distributions (Fig.~\ref{fig:cosmos_fainter}).  Both of these types of
systematic errors can, however, be accounted for by using pairs where one
object has a known spectroscopic redshift.

In principle the results from the close-pairs technique may be subject
to subtle biases related to the relationship between galaxy properties
and local environment.  For instance, if red sequence galaxies
preferentially appear in groups, then they will be over-represented in
a sample of close pairs.  Similarly, galaxies with boosted star
formation due to close interactions may also be over-represented.  On
the other hand, such galaxies may be relatively rare, and it is
worthwhile to remember that the number of pairs in a sample is a
strong function of the sample size itself, growing like $N^2-N$.
Another potential problem is that close pairs of objects may have
inaccurate photometry due to blending or poor background subtraction,
so it is important to apply a sensible lower limit for the pair
separations.

Our method of using angular associations of galaxies to constrain both
the redshifts and the redshift errors can be applied and extended in
various ways.  Particularly intriguing is the possibility of
incorporating information from angular associations directly into
photometric redshift codes.  A step in this direction has already been
taken by \citet{kovac09}, who modify the photometric redshift
probability distributions of objects that have near neighbors with
spectroscopic redshifts; here we simply note that this same idea may
be extended to neighbors with only photometric information.
Regardless of how the ideas discussed in this paper are used in
future, the close-pairs technique is straightforward to apply and
should prove to be a useful tool in analyzing data from redshift
surveys.

\acknowledgements

We are grateful to Simon White for stimulating discussions which
helped inspire this work, and to Marijn Franx, Pieter van Dokkum, and
Hendrik Hildebrandt for numerous useful comments.  We also thank
Hendrik for reducing the $u^*$-band data of the UDS field.  R.F.Q. is
supported by a NOVA Fellowship.

\end{document}